\documentclass[12pt]{JHEP3}

\title{Braneworld Cosmology in 
(Anti)--de Sitter Einstein--Gauss--Bonnet--Maxwell Gravity}

\author{James E. Lidsey \\
Astronomy Unit, School of Mathematical 
Sciences,  Queen Mary, \\ University of London,  
Mile End Road, LONDON, E1 4NS, U.K. \\
email: J.E.Lidsey@qmul.ac.uk
}
\author{
Shin'ichi Nojiri \\
Department of Applied Physics \\
National Defence Academy, 
Hashirimizu Yokosuka 239, JAPAN \\
email: nojiri@cc.nda.ac.jp
}
\author{
Sergei D. Odintsov \\
Tomsk Pedagogical University, 634041 Tomsk, RUSSIA \\
email: odintsov@ifug5.ugto.mx, odintsov@mail.tomsknet.ru
}
\abstract{
Braneworld cosmology for a domain 
wall embedded in the charged (Anti)-de Sitter-Schwarzschild
black hole of the five--dimensional Einstein-Gauss-Bonnet-Maxwell 
theory is considered. 
The effective Friedmann equation for 
the brane is derived by 
introducing the necessary surface counterterms 
required for a well-defined variational principle
in the Gauss--Bonnet theory and for the finiteness of the bulk space. 
The asymptotic dynamics of the brane cosmology is determined and 
it is found that solutions with vanishingly small spatial 
volume are unphysical. The finiteness of 
the 
bulk action is related to 
the vanishing of the effective cosmological constant 
on the brane. 
An analogy between the Friedmann equation 
and a generalized Cardy--Verlinde formula is drawn. 
}

\preprint{hep-th/0202198}

\keywords{dbr.ads.blh}

\begin{document}
\tolerance=5000

\def\pp{{\, \mid \hskip -1.5mm =}}
\def\cL{{\cal L}}
\def\be{\begin{equation}}
\def\ee{\end{equation}}
\def\bea{\begin{eqnarray}}
\def\eea{\end{eqnarray}}
\def\tr{{\rm tr}\, }
\def\nn{\nonumber \\}
\def\e{{\rm e}}
\def\D{{D \hskip -3mm /\,}}

\def\SEH{S_{\rm EH}}
\def\SGH{S_{\rm GH}}
\def\AdS5{{{\rm AdS}_5}}
\def\S4{{{\rm S}_4}}
\def\gfv{{g_{(5)}}}
\def\gfr{{g_{(4)}}}
\def\SC{{S_{\rm C}}}
\def\RH{{R_{\rm H}}}

\def\wlBox{\mbox{
\raisebox{0.1cm}{$\widetilde{\mbox{\raisebox{-0.1cm}\fbox{\ }}}$}}}
\def\htBox{\mbox{
\raisebox{0.1cm}{$\hat{\mbox{\raisebox{-0.1cm}{$\Box$}}}$}}}

\def\K{\left(k - (d-2) \e^{-2\lambda}\right)}

\def\double{\baselineskip 24pt \lineskip 10pt}
\renewcommand{\theequation}{\arabic{section}.\arabic{equation}}

\section{Introduction}

\setcounter{equation}{0}

\def\theequation{\thesection.\arabic{equation}}

The holographic principle in string/M--theory is formulated in terms 
of the 
AdS/CFT or dS/CFT correspondences \cite{adscft,dscft,hull}. 
These correspondences have
bridged the gap between previously distinct branches 
of high energy physics. In particular, 
it has become clear that the physics of higher--dimensional 
black holes is closely related to that of early universe 
cosmology. 
The possibility that our observable universe  
may be viewed as a domain wall or `brane' 
living on the boundary of a 
higher--dimensional black hole has recently been extensively 
discussed \cite{branerefs,RSII,othermoves,bv,ida}. 
In extensions to the second Randall--Sundrum model \cite{RSII}, 
for example, the brane is embedded in a five--dimensional
Schwarzschild--Anti de Sitter (SAdS) bulk and the mass of the 
black hole induces  a `dark radiation' term into the 
effective Friedmann equation on the brane \cite{bine}. This term 
alters the asymptotic behaviour of the cosmic dynamics at high 
energies \cite{maeda}. 
Thus far, attention has focused primarily on the case where 
the higher--dimensional bulk is represented by an 
AdS black hole (or more simply pure AdS space). 
However, the recently proposed dS/CFT correspondence 
provides motivation for considering 
the scenario where the bulk is (asymptotically) 
a de Sitter black hole \cite{dscft,hull}. 

In an attempt to place the braneworld scenario in a more 
string theoretic setting, a number of authors have considered 
the effects of introducing 
higher--derivative terms in the curvature 
\cite{HD,NOO,GBrefs,onlylocal,AM,cs,CD}.
Within the context of the AdS/CFT 
correspondence, higher curvature 
terms in the bulk theory can arise as next--to--leading order 
corrections in the $1/N$ 
expansion of the conformal field theory (CFT)
in the limit of strong coupling \cite{largeN}.
The Gauss Bonnet (GB) combination of curvature 
invariants is of particular interest to the braneworld cosmological 
scenario \cite{GBrefs,AM,cs,CD}. This combination appears naturally 
in the next--to--leading order term 
of the heterotic string effective action \cite{stringGB,bd}.
Although it contains higher--derivative gravitational 
terms in the metric, it is the unique combination in five 
dimensions that results in second--order 
field equations, albeit of a much more 
complicated form than that of standard Einstein 
gravity \cite{d86}. Furthermore, 
from the point of view of the Randall--Sundrum scenario \cite{RSII}, 
this property is crucial in ensuring a localization of 
gravity on the brane \cite{onlylocal}. 
The singular source associated with the brane 
manifests itself as a $\delta$--function in the energy--momentum 
tensor and this must be cancelled, at the level of the field 
equations, by second derivatives in the metric. 

In view of the above developments, therefore, 
the study of black holes 
and braneworlds in the GB theory is well motivated, 
both from the field theoretic and cosmological points of view.  
Recently, Cai found a class of topological black holes in 
$D$--dimensional Einstein--GB theory with a cosmological 
constant \cite{cai}, generalizing an earlier solution due to 
Boulware and Deser \cite{bd}. These solutions were further generalized 
to a class of charged SAdS and Schwarzschild--de Sitter
(SdS) black holes in the Einstein--GB--Maxwell 
theory, where a non--trivial electromagnetic field is present
\cite{cvetic}. 
In this paper, we investigate the dynamics of four--dimensional 
vacuum branes embedded in this class 
of five--dimensional (Anti)--de Sitter black holes.

The structure of the paper is as follows. We present the bulk black hole 
solutions  
and the general form of the braneworld Friedmann equation 
in Section 2. We proceed in Section 3 to derive the effective 
Friedmann equation on the brane for the case where the higher--dimensional 
spacetime is asymptotically AdS. We determine the asymptotic 
behaviour of the brane cosmologies in Section 4 for this case. 
In Section 5, we consider the corresponding scenario where 
the bulk space is asymptotically de Sitter.  
Section 6 contains  
a discussion on the (generalized) Cardy--Verlinde formulae 
\cite{EV,Cardy} in both Einstein and higher--derivative 
theories of gravity 
with a cosmological constant. We 
conclude with a discussion in Section 7.

\section{Bulk Black Hole Geometry and Brane Dynamics}

\setcounter{equation}{0}

\def\theequation{\thesection.\arabic{equation}}

The $(d+1)$--dimensional Einstein--GB--Maxwell bulk  
action has a matter sector 
consisting of a one--form gauge potential, ${A}_{\mu}$, 
with an antisymmetric electromagnetic field strength, 
${F}_{\mu\nu} \equiv \partial_\mu {A}_\nu - \partial_\nu 
{A}_\mu $, and a 
higher--order GB term in the 
gravitational sector of the theory.  
The action is given by\footnote{In this paper, 
upper case Latin indices run from $(A,B) =
(1,2,3)$ over the 
spatial sections of the world--volume of the brane, lower 
case Latin indices span the world--volume, $(i,j) =(0,1,2,3)$, 
lower case Greek indices span the bulk coordinates, 
$y$ is the component associated with the fifth dimension  
and a comma denotes partial differentiation.} 
\begin{eqnarray}
\label{vi}
{S}=\int d^{d+1} x \sqrt{-{g}}\left\{
{1 \over \kappa^2} {R} - \Lambda +
c\left( {R}^2 -4 {R}_{\mu\nu} {R}^{\mu\nu}
+ {R}_{\mu\nu\xi\sigma} {R}^{\mu\nu\xi\sigma}\right) 
\right. \nonumber \\
\left. -\frac{1}{4} {g}^{\mu\nu} {g}^{\rho\sigma} {F}_{\mu\rho} 
{F}_{\nu\sigma} \right\}  ,
\end{eqnarray}
where $c$ is an  arbitrary coupling constant, $\kappa^2$ parametrizes 
the $(d+1)$--dimensional Planck mass, the Riemann 
tensor, ${R}_{\mu\nu\xi\sigma}$, and its contractions are 
constructed from the metric, ${g}_{\mu\nu}$, and its 
derivatives, $g \equiv {\rm det} {g}_{\mu\nu}$ and $\Lambda$ represents 
the bulk cosmological constant.

By extremising the 
variations of the action (\ref{vi}) with 
respect to the metric tensor and gauge field, respectively, we obtain 
the field equations 
\bea
\label{R3}
0&=&{1 \over 2}{g}^{\mu\nu}\left\{c\left( {R}^2 
- 4 {R}_{\rho\sigma} {R}^{\rho\sigma}
+  {R}_{\rho\lambda\xi\sigma} {R}^{\rho\lambda\xi\sigma}\right)
+ {1 \over \kappa^2} {R} - \Lambda \right\} \nn
&& + c\left(-2{R}{R}^{\mu\nu} + 4 {R}^\mu_{\ \rho} 
{R}^{\nu\rho} 
+ 4 {R}^{\mu\rho\nu\sigma} {R}_{\rho\sigma} 
 - 2 {R}^{\mu\rho\sigma\tau}{R}^\nu_{\ \rho\sigma\tau} \right) \nn
&& - { 1 \over \kappa^2}{R}^{\mu\nu} + 
\frac{1}{2} \left( {F}^{\mu\sigma} {{F}^{\nu}}_{\sigma} 
-\frac{1}{4} {F}_{\sigma\rho} {F}^{\sigma\rho} {g}^{\mu\nu} 
\right)  \\
\label{emfield}
&& \partial_{\nu} \left( \sqrt{-g} F^{\mu\nu} \right) =0  .
\end{eqnarray}

We consider the case where the bulk spacetime 
corresponds to a static, hyper--spherically symmetric geometry with 
a line element given by 
\be
\label{GBiv}
ds^2 = - \e^{2\nu (r)} dt^2 + \e^{2\lambda (r)} dr^2 
+ r^2 \sum_{A,B=1}^{d-1} \tilde g_{AB} dx^A dx^B\ ,
\ee
where $\{ \nu (r) , \lambda (r) \}$ 
are functions of the radial coordinate, $r$, and the metric 
$\tilde g_{ij}$ is the metric of the $(d-1)$--dimensional Einstein manifold 
with a Ricci tensor defined by $\tilde R_{ij}=k g_{ij}$. 
The constant $k$ has values 
$k = \{ d-2 ,0 , -(d-2) \}$ for a 
$(d-1)$-dimensional unit sphere, a flat Euclidean space and 
a $(d-1)$-dimensional 
unit hyperboloid, respectively. 

Restricting to the five--dimensional case $(d=4)$, 
Eqs. (\ref{R3}) and (\ref{emfield}) admit the 
charged black hole solution \cite{cvetic}:
\bea
\label{sol1}
\e^{2\nu}&=&\e^{-2\lambda} \nn
&=& {1 \over 2c}\left\{ck + {r^2 \over 2\kappa^2 } \right.\nn
&& \left. \pm \sqrt{ {r^4 \over 4\kappa^4}
\left({4c\kappa^2 \over l^2} -1
\right)^2 - {2c\mu \over \kappa^2}\left({4c\kappa^2 \over l^2} -1
\right) - {2cQ^2 \over 3r^2 }} \right\}\ ,\\
\label{sol2}
F_{tr}&=&{Q \over r^3}\ ,
\eea
where the constants $\{ Q , \mu \}$ 
are related to the charge and gravitational mass of the 
black hole, respectively, and 
\be
\label{ll}
{1 \over l^2}\equiv{1 \over 4c\kappa^2 }\left(1\pm 
\sqrt{1 + {2c\Lambda \kappa^4 \over 3}}\right)  .
\ee
The constant, $l^2$, is 
determined by the Gauss--Bonnet coupling parameter, $c$, and the bulk 
cosmological constant, $\Lambda$. 
It 
corresponds to the length parameter of the asymptotically 
AdS space when $r$ is large. If $c\Lambda>0$, $l^2$ can be 
formally negative and the spacetime then becomes asymptotically de Sitter. 
(In this case, it represents a charged de Sitter space or the 
charged Nariai (SdS) black hole. See Ref. \cite{cvetic} 
for details).
In principle, the above solution (\ref{sol1})--(\ref{sol2}) 
for positive $l^2$
generalizes the 
well-known
Reissner-Nordstrom-AdS black hole solution to Einstein--GB--Maxwell gravity.
Moreover, it may also correspond to 
pure, charged AdS space and it has been observed that 
Hawking--Page phase transitions \cite{hawkingpage}
between the SAdS black hole and pure AdS 
space can be realized \cite{cvetic}.
We refer to the two 
solutions in Eq. (\ref{sol1}) as the `positive--' and 
`negative--branch' solutions, respectively. They
reduce to the class considered 
recently by Cai in the charge neutral limit $(Q=0)$ 
\cite{cai}.

We now proceed to consider the motion of a domain wall (three-brane) 
along a timelike geodesic 
of the five--dimensional, static background defined by Eqs.
(\ref{GBiv}) and (\ref{sol1}). 
The equation of motion of the brane is interpreted by an observer confined 
to the brane as an effective Friedmann equation describing the 
expansion or contraction of the universe. 
>From this Friedmann equation, we can deduce 
the energy and entropy of the matter in the brane universe. 
Specifically, we consider a brane action of the form: 
\begin{equation}
\label{braneaction}
S_{\rm br} = - \eta \int d^4 x \sqrt{-h}  ,
\end{equation}
where $\eta$ is a positive constant representing the 
tension associated with the brane and $h$ is the determinant of the 
boundary metric, $h_{ij}$, induced by the bulk metric, 
${g}_{\mu\nu}$.

We employ the method developed in Ref. \cite{NOO}
to derive the Friedmann equation. 
The metric  
(\ref{GBiv}) is rewritten   
by introducing new coordinates $ (y , \tau )$ and a scalar function
$A=A(y ,\tau )$ that satisfies the set of constraint equations:
\bea
\label{cc1}
&& l^2\e^{2A+2\lambda}A_{,y}^2 - \e^{-2\lambda} t_{,y}^2 = 1 \ , \nn
&& l^2\e^{2A+2\lambda}A_{,y}A_{,\tau}
- \e^{-2\lambda}t_{,y} t_{,\tau}= 0 \ ,\nn
&& l^2\e^{2A+2\lambda} A_{,\tau}^2 - \e^{-2\lambda} t_{,\tau}^2 
= -l^2\e^{2A}\ , 
\eea
where a comma denotes partial differentiation. 
When $\lambda =-\nu$, as in Eq. (\ref{sol1}), the metric (\ref{GBiv}) 
may then be written in the form 
\be
\label{metric1}
ds^2=dy^2 + \e^{2A(y,\tau)}\sum_{i,j=1}^4
\tilde g_{ij}dx^i dx^j  ,
\end{equation}
where $r= l\exp (A)$. The non--trivial Riemann components 
and its contractions for this metric are shown in the appendix. 
Since we are interested in the cosmological implications, we assume that the 
metric, $\tilde{g}_{ij}$, respects the same symmetries as the 
metric  of the Friedmann--Robertson--Walker (FRW) models, i.e., 
we assume that 
\begin{equation}
\label{worldvol}
\tilde g_{ij}dx^i dx^j \equiv l^2\left(-d \tau^2 
+ d\Omega_{k,3}^2\right)  ,
\ee
where $d\Omega_{k,3}^2$ is the metric of unit 
three--sphere for $k>0$, three--dimensional Euclidean space for 
$k=0$ and the unit three--hyperboloid for $k<0$. 
Thus, choosing a timelike coordinate, $\tilde t$, such that 
$d\tilde t \equiv l\e^A d\tau$, implies that  
the induced metric on the brane takes the FRW form: 
\be
\label{e3}
ds_{\rm brane}^2= -d \tilde t^2  + l^2\e^{2A}
d\Omega^2_{k,3}  .
\ee

It follows by solving Eqs. (\ref{cc1}) that 
\be
\label{e4}
H^2 = A_{,y}^2 - {\e^{-2\lambda}\e^{-2A} \over l^2}\ ,
\ee
where the Hubble parameter on the brane is 
defined by $H \equiv dA /d {\tilde t}$. 
Thus, for a vacuum brane that has no matter confined to it, the 
cosmic expansion (contraction) is determined once the functional 
forms of $\{ A , \lambda \}$ have been determined. 

\section{Brane Dynamics in Asymptotically AdS Space}

\setcounter{equation}{0}

\def\theequation{\thesection.\arabic{equation}}

In this Section we consider the brane equation of motion in an 
asymptotically AdS space, corresponding to $l^2 >0$. We defer the case 
$l^2<0$ to Section 5.

The functional form of $A_{,y}$ is deduced by deriving the 
surface counterterms 
in the Gauss--Bonnet theory (\ref{vi}) that generalize the Gibbons--Hawking 
term in Einstein gravity \cite{GH}. This has been done previously 
in Ref. \cite{cvetic} and we briefly summarize the approach here.
The reader is referred to \cite{cvetic} for details. 
Auxiliary fields, $F$,  $B_{\mu\nu}$ and $C_{\mu\nu\rho\sigma}$ are 
introduced and the Gauss--Bonnet sector of the action 
(\ref{vi}) is rewritten in terms of these variables 
by employing the field equations. Imposing a Dirichlet type of 
boundary condition allows appropriate boundary conditions to 
be
imposed on 
the scalar quantities. This in turn implies that the 
action can be expressed in terms of a bulk part and a surface term. 
The variational principle then becomes well defined if 
an appropriate boundary 
term is introduced. 
An important point is that the complete boundary action does 
not acquire terms from the form field, since the bulk action for 
the form field does not contain second-- or higher--order 
derivatives. Thus, we may extract the relevant equations 
from \cite{cvetic}, specialized to the Gauss--Bonnet 
interaction. In particular, the counterterm is 
identical to Eq. (130) of \cite{cvetic}: 
\bea
\label{Iiv}
S_b&=& \int d^4 x \sqrt{-h}\left[
4 c \nabla_\mu n^\mu  F -8c\left(n_\mu n_\nu
\nabla_\sigma n^\sigma
 + \nabla_\mu n_\nu \right)  B^{\mu\nu} \right. \nn \\ 
&& \left. + 8 c n_\mu n_\nu \nabla_\tau n_\sigma
 C^{\mu\tau\nu\sigma}
+ {2 \over \kappa^2}\nabla_\mu n^\mu 
- \eta \right] \ ,
\eea
where $n^{\mu}$ is the unit vector perpendicular to the 
boundary and we have also included the Gibbons--Hawking term and 
the counterterm (\ref{braneaction}) arising from the vacuum  energy 
on the brane.  
Extremizing the variation over $F$ of the full 
action\footnote{A factor of two arises because two bulk 
spacetimes with a common boundary are being considered.
In \cite{RSII}, two copies of the AdS spaces are glued by a brane 
by imposing the $Z_2$-symmetry. In deriving the
corresponding Friedmann 
equation in 
\cite{NOO}, essentially only one side was considered. 
In the situation that we are considering above, there 
are two copies of the AdS 
space as in \cite{RSII}, and we therefore need twice the surface 
counterterms.}
$(S+2S_b)|_{y=y_0}$ then implies that 
\bea
\label{viii}
0&=& 16cR_{,y} -8c \left(4R_{yy,y} + R_{i\ ,y}^{\ i}\right)
+8c  C_{yiy\ ,y}^{\ \ \ i} \nn
&& + \left( 48 c F -192c  B_{yy}
 + c \left(56 R_{yy} - 8 R_{i}^{\ i}\right)+{24 \over \kappa^2}
\right)A_{,y} + 4\eta  .  
\eea
It is important to emphasize that the boundary terms are derived 
from the internal consistency of the theory and are not 
simply introduced in an {\em ad hoc} fashion. 

When the metric has the form (\ref{GBiv}{}) with $\lambda=-\nu$, 
the Ricci scalar and non--trivial components 
of the Ricci tensor are given by 
\def\BB{\left({1 \over 4\epsilon l^2}\right)}
\def\CC{{\left(4\epsilon - 1\right)^2 \over 4\kappa^4}}
\def\DD{\left(2\epsilon{\tilde{\mu}} \left(4\epsilon  - 1\right)\right)}
\def\EE{(\tilde{Q}^2)}
\bea
\label{RA1}
\lefteqn{R= - \left(\e^{2\nu}\right)'' - {6 \over r}
\left(\e^{2\nu}\right)' - {6 \over r^2}\e^{2\nu}
+ {3k \over r^2}} \nn
\lefteqn{= {5 \over \epsilon l^2} \mp
{1 \over 2cr^6}\left\{2\tilde{Q}^4 + 18\epsilon {\tilde{\mu}} (4\epsilon -1)
\tilde{Q}^2 r^2 + 6\DD^2 r^4 \right.} \nn
&& \left. - {31 \left(4\epsilon - 1\right)^2 \over
4\kappa^4} \tilde{Q}^2 r^6 - {15 \left(4\epsilon - 1\right)^3
\epsilon{\tilde{\mu}}
\over \kappa^4} r^8
+ 20 \left(\CC\right)^2 r^{12}\right\} \nn
&& \times \left\{ -\DD - {\tilde{Q}^2 \over r^2} + \CC r^4
\right\}^{-{3 \over 2}} 
\end{eqnarray}
and
\begin{eqnarray}
\label{RA2}
\lefteqn{R_{yy}= \e^{2\nu} R_{rr}
= - {1 \over 2} \left(\e^{2\nu}\right)'' - {3 \over 2r}
\left(\e^{2\nu}\right)'} \nn
\lefteqn{= - {1 \over \epsilon l^2} \mp {1 \over 2cr^6}
\left\{-\tilde{Q}^4 - {4\left(4\epsilon - 1\right)^2 \over \kappa^4}
\tilde{Q}^2 r^6 - {6\left(4\epsilon - 1\right)^3\epsilon{\tilde{\mu}} 
\over \kappa^4} r^8 \right. }\nn
&& \left. + 8 \left(\CC\right)^2 r^{12}\right\}
\times \left\{ - 2\epsilon{\tilde{\mu}} \left(4\epsilon  - 1\right)
 - {\tilde{Q}^2 \over r^2} + \CC r^4 \right\}^{-{3 \over 2}} \ .
\eea
respectively, where a prime denotes differentiation with 
respect to $r$, we have defined the rescaled 
parameters 
\be
\label{E1}
\epsilon \equiv{c\kappa^2 \over l^2} , \quad \tilde{\mu} \equiv 
\frac{l^2 \mu}{\kappa^4} , \quad \tilde{Q}^2 \equiv \frac{2c}{3} Q^2
\ee
to simplify the notation 
and
the coordinates $y$ and $r$ are related to each other
by $dy=\e^{-\nu}dr$. 
Substituting Eqs. (\ref{RA1}) and (\ref{RA2}) into Eq. 
(\ref{viii}) then yields a relationship between $A_{,y}$ and the 
parameters of the bulk solution:
\bea
\label{RA3}
\lefteqn{0=4 \eta \pm \left.\left.{12 \over r^7} \right\{ 3\tilde{Q}^4
+ 16 \epsilon^2 {\tilde\mu}^2 \left(4\epsilon - 1\right)^2 r^4 
+ 6\tilde{Q}^2 r^2 \right(2\epsilon \tilde\mu 
\left(4\epsilon - 1\right) } \nn
&& \left.\left. + {\left(4\epsilon - 1\right)^2 \over 4\kappa^4}r^4
\right)\right\} 
\times \left\{ - 2c{\tilde{\mu}}(4\epsilon - 1) - {\tilde{Q}^2 \over r^2}
+ \CC r^4 \right\}^{-{3 \over 2}} \nn
&& \times \left[ {k \over 2} + \BB r^2
\pm {1 \over 2c}\left\{ - 2\epsilon {\tilde{\mu}} (4\epsilon - 1)
 - {\tilde{Q}^2 \over r^2} + \CC r^4
\right\}^{1 \over 2}\right]^{1 \over 2} \nn
&& + \left[-24c\left\{ {3 \over \epsilon l^2} \pm {1 \over 2cr^6}
\left\{6\tilde{Q}^4  + 3\tilde{Q}^2 
\left(10\epsilon{\tilde{\mu}} (4\epsilon - 1) r^2
- {3\left(4\epsilon - 1\right)^2 \over 4\kappa^4} r^6\right)
\right.\right.\right. \nn
&& \left. + 2r^4 \left(5\DD^2 + 6 \left(\CC\right)^2 r^8
 - {9\left(4\epsilon - 1\right)^3\epsilon
 {\tilde{\mu}} \over 2\kappa^4}
r^4\right)\right\} \nn
&& \left.\left. \times \left\{ -2\epsilon{\tilde{\mu}} ( 4\epsilon - 1)
 - {\tilde{Q}^2 \over r^2} + \CC r^4 \right\}^{-{3 \over 2}}\right\}
+ {24 \over \kappa^2}\right]A_{,y} \ .
\eea

It now only remains to identify 
the scale factor of the metric on the brane, 
$a(\tilde t)\equiv l\e^A$, with the radial 
coordinate, $r$, 
of the brane in the bulk metric (\ref{GBiv}):
\be
\label{GBvxvii}
a=r(\tilde t)=l\e^{A(\tilde t)}\ .
\ee
By combining Eqs. (\ref{e4}), (\ref{RA3}) and (\ref{GBvxvii}), 
we arrive at 
the Friedmann equation describing the motion of the 3-brane 
in the bulk spacetime. We find it convenient 
to express this equation in the compact form 
\begin{equation}
\label{RA5}
H^2 = \frac{{\cal{G}}^2}{{\cal{H}}^2} 
- {X (a) \over a^2}   ,
\end{equation}
where 
\begin{eqnarray}
\label{defG}
{\cal{G}}&=&  4 \eta \pm \left. 
\frac{12X^{1/2}}{Y^{3/2} a^7} \right\{ 3\tilde{Q}^4
+ 16 \epsilon^2 {\tilde\mu}^2 \left(4\epsilon - 1\right)^2 a^4 \nn
&& \left. + 6\tilde{Q}^2 a^2 \left(2\epsilon \tilde\mu 
\left(4\epsilon - 1\right) 
+ {\left(4\epsilon - 1\right)^2 \over 4\kappa^4}a^4
\right)\right\} 
\eea
and 
\begin{eqnarray}
\label{defH}
{\cal{H}}  = -\frac{48}{\kappa^2}
\mp {12  \over Y^{3/2} a^6}
\left\{6\tilde{Q}^4 +3\tilde{Q}^2 
\left(10\epsilon {\tilde{\mu}} (4\epsilon - 1) a^2
 - {3\left(4\epsilon - 1\right)^2 \over 4\kappa^4} a^6\right)
\right.    \nn \\
\left.  + 2a^4 \left( 5\DD^2 + 6 \left(\CC\right)^2 a^8
\right. \right.  \nonumber \\
\left. \left. - {9\left( 4\epsilon - 1\right)^3 \epsilon {\tilde{\mu}} 
 \over 2\kappa^4} a^4\right) \right\}  
\eea
and we have identified the two important 
functions of the scale factor: 
\begin{eqnarray}
\label{X} 
X \equiv \frac{k}{2} + \frac{a^2}{4\epsilon l^2} \pm 
{Y^{1/2} \over 2c}
\\
\label{Y}
Y \equiv -2\epsilon {\tilde{\mu}} (4\epsilon -1 )
 -\frac{\tilde{Q}^2}{a^2} 
+\frac{(4\epsilon -1 )^2}{4\kappa^4} a^4  .
\end{eqnarray}

In the following Section, we proceed to analyze 
the qualitative early-- and 
late--time behaviour of the cosmic dynamics on the 
brane. 

\section{Qualitative Dynamics of the Brane Cosmology}

\setcounter{equation}{0}

\def\theequation{\thesection.\arabic{equation}}

The simplest case to consider arises 
when the black hole has vanishing 
mass and charge. In this case, the Friedmann 
equation (\ref{RA5}) simplifies to
\begin{equation}
\label{simfri}
H^2 = \frac{\eta^2\kappa^4}{36} \left[ 2 \pm 
3|4\epsilon -1| \right]^{-2} 
-\frac{k}{2a^2} -\frac{1}{4\epsilon l^2} \left( 1\pm 
|4\epsilon -1 | \right)  .
\end{equation}
We should note that the $\pm$ signs in (\ref{simfri}) 
correspond to the two solutions in (\ref{ll}). 
Formally, this is equivalent to the standard Friedmann equation, 
containing a cosmological constant term and a curvature term. 
The three parameters, $\{ \eta , c , l^2 \}$, 
in Eq. (\ref{simfri}) are related by requiring that 
the leading--order divergence in 
the bulk AdS action is cancelled when the brane moves to 
infinity in the asymptotic AdS space.
The divergence is cancelled if the brane tension satisfies \cite{NOO}
\begin{equation}
\label{cancel}  
{\eta \over l}={72 c \over l^4} - 
{6 \over \kappa^2 l^2}   .
\end{equation}
When no higher--order curvature invariants are 
included in the action (\ref{vi}), 
Eq. (\ref{cancel}) reduces to the fine--tuning 
condition imposed by Randall and Sundrum 
for the vacuum brane to respect four--dimensional 
Poincare invariance \cite{RSII}.  
Once the brane tension has been specified by Eq. (\ref{cancel}), 
there are no {\em a priori} 
restrictions on the remaining two parameters, and 
in principle,
an appropriate choice can 
then be made to yield an effective positive cosmological 
constant on the brane. Consequently, the
brane can undergoe a de Sitter phase of expansion 
(inflation).  

An interesting special case arises for $l^2 =4c \kappa^2$ $(\epsilon =1/4 )$, 
where the two branches of the bulk solution (\ref{sol1}) 
coincide.
Eq. (\ref{cancel}) then implies that $\eta l \kappa^2 =12$ and 
when this expression is substituted into Eq. (\ref{simfri}), 
the effective cosmological constant {\em vanishes}. In this model, 
therefore, the vanishing of the cosmological constant 
is associated with the cancellation of 
the leading--order divergence of the action. 
In the limit of 
small Gauss--Bonnet coupling, $\epsilon \ll 1$, the brane 
tension also cancels the contributions from the five--dimensional cosmological 
constant when the negative--branch solution is considered and 
Eq. (\ref{cancel}) is satisfied.  
Unless otherwise stated, we consider 
this branch in what follows\footnote{The bulk 
solution for the positive--branch 
has been shown to be unstable \cite{bd}.}.  

It is of interest to determine how the expansion of the braneworld 
is affected by the 
charge and mass associated with the bulk black hole. 
The advantage of the parametrization (\ref{defG})--(\ref{Y}) 
is that it separates 
the Friedmann equation (\ref{RA5}) into two sectors 
that 
both admit a physical interpretation. 
The first term is determined by the ratio, ${\cal{G}}^2/{\cal{H}}^2$,
and the numerator, Eq. (\ref{defG}), contains 
the brane tension, $\eta$, and some further
terms that depend on the 
parameters of the bulk black hole. 
In the limit of small Guass--Bonnet coupling, $\epsilon \ll 1$, 
and vanishing charge, 
we find that ${\cal{G}} =4 \eta +{\cal{O}} (\epsilon^2 )$. 
In effect, therefore, 
this term parametrizes the vacuum 
energy of the brane. By analogy with the standard, four--dimensional 
Friedmann equation, the denominator, Eq. (\ref{defH}), then
plays the role 
of the effective Planck mass. This parameter is time--dependent, 
in general, 
due to the non--trivial black hole mass and charge. 
To lowest--order in $\epsilon$, it also 
reduces to a constant,  
${\cal{H}} = 24 ( 1- 12 \epsilon )/\kappa^2 
+{\cal{O}} (\epsilon^2 )$, when the charge vanishes. 

The second sector of the 
Friedmann equation, $X/a^2$, may be interpreted as a generalized 
curvature term. This sector depends 
directly on the spatial curvature of the world--volume, $k$,
but also acquires corrections due to deviations of the 
bulk geometry from pure ${\rm AdS}$ space. 
Indeed, Eq. (\ref{X}) reduces to a constant 
value in the limit where the first term on the right--hand side of this 
equation is dominant. 

When Eq. (\ref{cancel}) is satisfied and the black hole charge vanishes, 
the limit of the negative--branch Friedmann equation (\ref{RA5})
as $\epsilon \rightarrow 
0$ is given by 
\begin{equation}
\label{smallf}
H^2 = -\frac{k}{2a^2} +\frac{\mu}{a^4} -\frac{2c \kappa^2 \mu^2}{a^8}
+{\cal{O}}( c^2)   ,
\end{equation}
where we have returned to the original parameters (\ref{E1}).
As in the massless case considered above, Eq. (\ref{cancel}) ensures  
that the cosmological terms are cancelled. At this order, the leading 
terms in Eq. (\ref{smallf}) correspond to those of the 
second Randall--Sundrum model \cite{RSII}, as expected. 
The term 
due to the Gauss--Bonnet sector has a strong dependence on the scale factor
and is rapidly redshifted by the cosmic expansion. 
Introducing a bulk charge in this limit results in a complicated expression 
for the Friedmann equation, 
where all but one of the new terms that arise do so 
with a linear dependence on $\epsilon$. 
The remaining 
term arises from the expansion of Eq. (\ref{X}) and plays the role 
of a shear or anisotropic stress. It corresponds to the 
contribution that arises when the brane is embedded  
in the Reissner--Nordstrom--AdS black hole
bulk space \cite{bv,csaki}. 

More generally, 
the functions (\ref{X}) and (\ref{Y}) play a crucial 
role in determining the asymptotic behaviour of the braneworld
and, in particular, 
they must both remain semi--positive definite 
if physical solutions to Eq. (\ref{RA5}) are to exist. (These functions 
are not constrained in this way in the special case 
where $\tilde{Q}^2 =\tilde{\mu} =0$, since Eq. (\ref{defG}) 
then depends only on the brane tension).
We may deduce immediately, therefore, that the 
effective curvature term in the Friedmann equation (\ref{RA5}) always 
acts to {\em reduce} the expansion rate of the brane if the 
black hole has a mass and/or a charge.  
An alternative interpretation is that the brane expands as 
if its effective curvature 
is positive and this is independent of its specific 
curvature, $k$. 
An important consequence of this property is that the brane 
may recollapse before the effective cosmological constant on the 
brane is able to dominate the dynamics\footnote{We are assuming implicitly 
in this discussion that the effective cosmological constant 
on the brane is positive.}. This is true 
for all spatially open, flat and closed branes
and therefore differs from the second Randall--Sundrum 
scenario based on the SAdS bulk space, 
where recollapse is only possible in the positively curved model
\cite{RSII}. 

In determining the asymptotic behaviour, 
we first consider 
the large world--volume limit, $a \rightarrow \infty$. 
It is straightforward to verify that both 
the mass and charge become negligible in 
Eqs. (\ref{X}) and (\ref{Y}) as the scale factor diverges. 
However, to ensure that the function (\ref{X}) remains positive in this limit, 
we require that $1\pm |4 \epsilon -1 | > 0$ when $\epsilon >0$ and 
$1\pm |4 \epsilon -1 | < 0$ when $\epsilon <0$. 
There is no corresponding limit on the value of the black hole mass. 
Moreover, the leading--order term in Eq. 
(\ref{defG}) redshifts as the eighth (sixth) power 
of the scale factor when the charge is trivial (non--trivial). 
Thus, this function tends to a constant, ${\cal{G}} \rightarrow 4 \eta$, 
as in the massless, neutral solution. Similarly, the 
dominant term in Eq. (\ref{defH}) is the term corresponding to 
$\tilde{\mu} =\tilde{Q} =0$. The 
asymptotic 
behaviour in the region of parameter space where the 
scale factor can become arbitrarily large 
is therefore given by Eq. (\ref{simfri}). 
This behaviour can be understood from a physical point of view, 
since an increase in the spatial volume 
of the brane corresponds to the brane 
being located at 
greater distances from the event horizon of the black hole. 

An ambiguity arises, however, when considering the 
opposite limit, where the scale factor becomes small. 
Let us first consider the case 
where the black hole is massive but has vanishing charge.
The range of the Gauss--Bonnet parameter, $\epsilon$, 
is important in this case. 
In the region of parameter space $0 < \epsilon < 1/4$, 
Eq. (\ref{Y}) is positive--definite 
for all allowed values of the scale factor. 
Further motivation for considering 
this range of values for $\epsilon$ arises from the asymptotic form 
of the bulk solution (\ref{sol1}) as $r$ diverges. In this 
limit, the metric reduces to the standard 
Reissner--Nordstrom--AdS black hole with a real charge \cite{cvetic}. 
On the other hand,  
Eq. (\ref{X}) implies that for the non--positively curved brane, $k \le 0$, 
it is necessary to take the positive--branch if $a$ is to   
become arbitrarily small. In the negatively curved model, 
the further constraint, $| \epsilon (4\epsilon -1 )| > 2c^2 /\tilde{\mu}$, 
must also be satisfied. There is no constraint for the positively 
curved brane on the positive branch solution, but the corresponding 
limit $|Y|^{1/2} < 2c$ must be satisfied for the negative--branch. 

Nevertheless, even when the above necessary conditions 
are satisfied, we find that 
the ratio
\begin{equation}
\label{ratio}
\frac{{\cal{G}}^2}{{\cal{H}}^2} \rightarrow \frac{4X}{25a^2}
\end{equation}
in the limit $a \rightarrow 0$.
Substituting this limit into the Friedmann equation 
(\ref{RA5})
then implies that $H^2<0$. Although this is indicative of 
unphysical behaviour, it 
is interesting because it implies that when 
$0< \epsilon < 1/4$, 
the scale factor of the vacuum brane 
is bounded from below 
and consequently can not approach zero in a continuous fashion. 
It is interesting that this 
conclusion holds for arbitrary values of the black hole mass. 

In the remaining region of parameter space, 
$\epsilon (4 \epsilon -1 ) >0$, 
it follows that 
$Y \rightarrow 0$ at some finite value of the 
scale factor as the 
spatial volume  
decreases, assuming implicitly that the black hole mass parameter 
${\tilde{\mu}} >0$. 
Although the Hubble parameter
remains finite as $Y \rightarrow 
0$, 
the Ricci curvature of the world--volume diverges and this 
corresponds to a singular initial state of finite volume. 
It is difficult to motivate such behaviour from a physical 
point of view. 

To summarize, thus far, there is a lower limit 
to the spatial volume associated 
with physical cosmological solutions 
on the brane world--volume. 
This conclusion holds 
for all values of $\epsilon$, and
hence the Gauss--Bonnet 
coupling, $c$. It is also 
independent of the value of the 
bulk cosmological constant (modulo the restriction 
imposed in this Section that $l^2 >0$). 
This implies that the vacuum 
brane can not expand from a big bang initial (singular) state 
of vanishing spatial volume. 

When the black hole charge is non--trivial, 
the functional 
forms of Eqs. (\ref{X}) and (\ref{Y}) differ in the limit
of small scale factor. In particular,
they are dominated by the term proportional to 
$\tilde{Q}^2/a^2$, indicating that a necessary condition 
for the brane to expand from an arbitrarily 
small volume is that $\tilde{Q}^2 <0$. 
Since the charge of the black hole should be real, this implies that 
$\epsilon < 0$ ($c< 0$). In this case,  it 
is possible to find appropriate values of the 
charge to ensure that $Y$ is positive--definite. 
On the other hand, since Eq. (\ref{X})  becomes dominated by the third term
on the right hand side, we must choose   
the positive branch in this limit. 
However, in this region of parameter space, 
we deduce that 
\begin{equation}
\label{Qratio}
\frac{{\cal{G}}^2}{{\cal{H}}^2} \rightarrow \frac{X}{4a^2}
\end{equation}
and it follows, after 
substitution into Eq. (\ref{RA5}), 
that the scalar factor of the world--volume is once more bounded from below, 
as in the charge neutral case. 

We conclude, therefore, that 
the charge and mass of the bulk black hole have significant 
effects on the asymptotic behaviour of the brane cosmology 
at small spatial volumes, but become negligible in the region of 
parameter space where expansion is unbounded from above. 

Before concluding this 
Section, we consider a second special case where the restriction on the 
positivity of Eq.
(\ref{X}) need not be imposed. This arises  when the tension
of the brane formally vanishes, $\eta \rightarrow 0$.
Although the interpretation
of a vacuum brane with vanishing tension is
unclear (note that 
the effective tension does not vanish), the function $X$ in this 
case appears linearly in the Friedmann equation (\ref{RA5}).
Moreover, it follows from Eq. (\ref{cancel}) that $\epsilon =
1/12$ when the brane tension
vanishes and it is interesting that this is precisely the critical value for
the thermodynamical entropy 
associated with the bulk black hole spacetime to also vanish
\cite{cvetic}. This value of
$\epsilon$ ensures that Eq. (\ref{Y}) is always positive
when the black hole is neutral and we may therefore
consider the small volume behaviour of the brane\footnote{We do not
consider the charged case, since this requires $\epsilon <0$ for
consistency.}.
The limit of the 
Friedmann equation (\ref{RA5}) is then 
deduced from Eq. (\ref{ratio}): 
\bea
\label{special}
H^2 &=& {X \over a^2}\left(-{21 \over 25} \mp {4a^2 
\over 25\kappa^2 \sqrt{\tilde \mu}}\right) \nn
&=& -{21 \over 25a^2}\left({k \over 2}\mp {\sqrt{\tilde\mu} 
\over 6c}\right) - {63 \over 25 l^2} \\
&& \mp \left({k \over 2}\mp {\sqrt{\tilde\mu} \over 6c}\right) 
{4 \over 25 \kappa^2 \sqrt{\tilde \mu}}
+ {\cal O}\left(a^2\right)  .
\eea

Eq. (\ref{special}) implies that 
the brane behaves as if it were dominated by 
its spatial curvature, in the sense that $H \propto a^{-1}$. (There is a 
subdominant negative cosmological constant). 
The specific curvature of the 
spatial hypersurfaces of the induced metric of the brane is 
uniquely specified by the sign of $k$.  
However, the brane acts 
as if it had {\em arbitrary} curvature, because 
the sign of the first term on the right hand side of 
Eq. (\ref{special}) is not specified {\em a priori}. 
It is determined by the 
mass of the black hole and the Gauss--Bonnet coupling. 

This concludes the discussion on brane dynamics in asymptotically 
AdS bulk spaces. In the next Section, we determine the 
form of the Friedmann equation when the bulk space is asymptotically 
de Sitter.

\section{Brane Dynamics in Asymptotically dS Space}

\setcounter{equation}{0}

\def\theequation{\thesection.\arabic{equation}}

In the case where $l^2$, as defined in Eq. (\ref{ll}), 
is formally negative,  
the higher--dimensional bulk 
spacetime is asymptotically de Sitter space. In order to analyse 
this region of parameter space, it proves convenient to 
define 
\be
\label{dS1}
\bar l^2\equiv - l^2 > 0\ .
\ee
If we then consider a metric of the form (\ref{metric1}), 
where 
\be
\label{worldvoldS}
\bar g_{ij}dx^i dx^j \equiv \bar l^2\left(-d \tau^2 
+ d\Omega_{k,3}^2\right)\ ,
\ee
is introduced instead of Eq. (\ref{worldvol}), we find that 
the effective Hubble parameter on the brane is given by 
\be
\label{e4dS}
H^2 = A_{,y}^2 - {\e^{-2\lambda}\e^{-2A} \over \bar l^2}\ .
\ee

We may now follow the same argument as that presented in Section 
3 to deduce the form of $A_{,y}$. Omitting 
the details,  we find
after defining the parameters:
\be
\label{E1dS}
\bar \epsilon \equiv{c\kappa^2 \over \bar l^2}, \quad 
\bar{\mu} \equiv \frac{\mu \bar{l}^2}{\kappa^4} , \quad
\bar{Q}^2 =\frac{2c}{3} Q^2
\ee
that the Friedmann equation 
has the form 
\def\tBB{\left({1 \over 4\bar\epsilon \bar l^2}\right)}
\def\tCC{{\left(4\bar\epsilon + 1\right)^2 \over 4\kappa^4}}
\def\tDD{\left(-2\bar\epsilon\bar{\mu} 
\left(4\bar\epsilon  + 1\right)\right)}
\def\tEE{\left( \bar{Q}^2 \right)}
\begin{equation}
\label{RA5dS}
H^2 = \frac{\bar{\cal{G}}^2}{\bar{\cal{H}}^2}
-\frac{\bar{X}}{a^2}   ,
\end{equation}
where
\begin{eqnarray}
\label{barcalG}
\bar{\cal{G}}&=& 4 \eta \pm \left. 
\frac{12\bar X^{1/2}}{\bar Y^{3/2} a^7} \right\{ 3{\bar Q}^4
+ 16 \bar\epsilon^2 {\bar\mu}^2 
\left(4\bar\epsilon + 1\right)^2 a^4 \nn
&& \left. + 6{\bar Q}^2 a^2 \left(2\bar\epsilon \bar\mu 
\left(4\bar\epsilon + 1\right) 
+ {\left(4\bar\epsilon + 1\right)^2 \over 4\kappa^4}a^4
\right)\right\} 
\end{eqnarray}
\begin{eqnarray}
\label{barcalH}
\bar{\cal{H}} \equiv
-\frac{48}{\kappa^2}
\mp \frac{12}{\bar Y^{3/2} a^6}
\left\{ 6\bar{Q}^4 +3\bar{Q}^2
\left(-10\bar\epsilon \bar{\mu} (4\bar\epsilon + 1) a^2
 - {3\left(4\bar\epsilon + 1\right)^2 \over 4\kappa^4} a^6\right)
\right.    \nn
 \left.    + 2a^4 
\left(5\tDD^2 + 6 \left(\tCC\right)^2 a^8 \right. \right. \nn
 \left. \left. 
 + {9\left(4\bar\epsilon + 1\right)^3 \bar\epsilon \bar{\mu} 
 \over 2\kappa^2}
a^4\right) \right\}  
\end{eqnarray}
and
\begin{eqnarray}
\label{XdS}
\bar{X} \equiv \frac{k}{2} +\frac{a^2}{4\bar{\epsilon} \bar{l}^2} \pm 
{\bar{Y}^{1/2} \over 2c} \\
\label{YdS}
\bar Y \equiv 2\bar\epsilon \bar{\mu} (4\bar\epsilon +1 )
 -\frac{\bar{Q}^2}{a^2} 
+\frac{(4\bar\epsilon +1 )^2}{4\kappa^4} a^4  .
\end{eqnarray}

The dependence on the 
scale factor, $a$,
of the Friedmann equation (\ref{RA5dS})  is identical to that 
given in Eq. (\ref{RA5}). Thus, in general, Eqs. (\ref{XdS}) and (\ref{YdS}) 
must remain positive definite  
and the asymptotic limit of the brane in the large
volume limit is qualitatively similar to the corresponding model where
the bulk space is asymptotically AdS ($l^2>0$). The region of parameter space 
where such a limit is consistent is given by 
$1\pm | 4\bar{\epsilon}+1 | >0$ if $\bar{\epsilon} >0$ and 
$1\pm | 4\bar{\epsilon}+1 | <0$ if $\bar{\epsilon} <0$. 
Finally, similar conclusions to those of previous Section  
are drawn regarding the restrictions on the small volume limit. 

The physical interpretation 
of the functions (\ref{barcalG}), (\ref{barcalH}) and (\ref{XdS}) 
may also be made along similar lines to that 
discussed in Section 4. In the next Section, we 
consider the physical interpretation of the 
Friedmann equation (\ref{RA5}) in more detail  
and investigate whether an analogy 
can be made between this equation and a generalized Cardy--Verlinde formula 
\cite{EV,Cardy}. 

\section{Cosmological Entropy and Generalized 
Cardy-Verlinde Formula}

\setcounter{equation}{0}

\def\theequation{\thesection.\arabic{equation}}

As we have seen in the previous Sections, the GB term in action (\ref{vi})
results in a complicated form for the 
Friedmann equation (\ref{RA5}). Recently,
Verlinde \cite{EV} drew an interesting 
analogy between the FRW equations of a standard, closed, radiation--dominated 
universe 
and the two-dimensional entropy formula due to Cardy \cite{Cardy}.
In this Section, 
we will extend this analogy to the FRW equations under consideration.

Let us first investigate this analogy in Einstein gravity for 
the usual 
$(n+1)$-dimensional FRW Universe with a metric
\be
\label{F1}
ds^2=-d\tau^2 + a^2(\tau) g_{ij}dx^i dx^j  ,
\ee
where the $n$-dimensional spatial hypersurfaces  with negative, zero or 
positive curvature are  parametrized by $k=-1$, $0$, $1$, 
respectively\footnote{
In our discussion of cosmology based on 
Einstein gravity, we parametrize the curvature of 
the spatial hypersurfaces 
in terms of 
$k=-1$, $0$, $1$ since direct comparison
with the standard cosmological equations is then straightforward.}. 
We limit our discussion mainly to that of the closed 
universe ($k=1$), with a spatial volume defined 
by $V=a^n \int d^n x
\sqrt{g}$. 
The standard FRW 
equations may then be written as
\bea
\label{F2}
H^2&=&{16\pi G \over n(n-1)}\rho - {k \over a^2} \ ,\nn
\dot H&=&- {8\pi G \over (n-1)}\left(\rho + p \right) + 
{k \over a^2}  ,
\eea
where $\rho=\rho_m + {\Lambda \over 8\pi G}$, $p=p_m 
- {\Lambda \over 8\pi G}$, $\Lambda$ is a cosmological 
constant  and $\rho_m$ and $p_m$ are 
the energy density and pressure of the 
matter contributions. The energy conservation equation is 
\be
\label{F3}
\dot\rho + n(\rho + p){\dot a \over a}=0 
\ee
and, for a perfect fluid matter source with equation of 
state $p_m=\omega\rho_m$ $(\omega = {\rm constant})$, 
Eq. (\ref{F3}) is solved as:
\be
\label{F4}
\rho=\rho_0a^{-n(1+\omega)} + {\Lambda \over 8\pi G}\ .
\ee

The definitions for the 
Hubble, Bekenstein and Bekenstein-Hawking 
entropies are \cite{EV}:
\be
\label{F5}
S_H=(n-1){HV \over 4G}\ ,\quad 
S_{BH}=(n-1){V \over 4Ga}\ ,\quad
S_B={2\pi a \over n}E\ ,
\ee
where the total energy, $E$, is defined as $E=\rho V$ and
contains the contribution from the cosmological constant 
term. This differs from that of the standard case, where 
the definitions of the entropies $S_{BH}$ and $S_B$ 
may differ slightly in their coefficients. This is 
specific to the presence of a cosmological constant 
\cite{wang,ryan}. 

By employing the  above definitions (\ref{F5}), one can easily rewrite 
the FRW equations (\ref{F2}) as a 
cosmological Cardy-Verlinde (CV) formula:
\bea
\label{F6}
S_H&=&{2\pi \over n}a \sqrt{E_{BH}\left(2E - k E_{BH}\right)}\ , \nn
kE_{BH}&=&n\left(E + pV - T_H S_H\right)\ ,
\eea
where the energy and Hawking temperature of the black hole are defined as
\be
\label{F7}
E_{BH}=n(n-1){V \over 8\pi G a^2}\ ,\quad 
T_H=-{\dot H \over 2\pi H}\ .
\ee
and we have separated the energy into a matter part and a cosmological 
constant part, i.e., 
$E=E_m + E_{\rm cosm}$, where 
$E_{\rm cosm}={\Lambda \over 8\pi G}V$. 
This is simply a way to rewrite the FRW equations in 
a form that resembles the equation defining 
the entropy of a two--dimensional CFT. 
However, the following remark is in order: 
the presence of cosmological constant may change 
some of the coefficients in Eq. (\ref{F6}) and this depends on 
precisely how the separation between the strongly and 
weakly interacting gravitational phases is made (compare with 
\cite{wang,ryan}). In any case, the energy associated with 
the cosmological 
constant term is hidden in the expression for $E$, Eq.  (\ref{F6}). 

Eq. (\ref{F5}) may also be rewritten in another form:
\be
\label{F8}
S_H^2 = S_{BH} \left(2S_B - k S_{BH}\right)\ .
\ee
Since the definition of $S_B$ normally contains only matter 
contributions, it is reasonable to 
define $S_B \equiv S_B^m + S_B^{\rm cosm}$, where the 
entropy associated with the cosmological 
constant is given by 
\be
\label{F9}
S_B^{\rm cosm}= {aV\Lambda \over 4nG}\ .
\ee
The appearance of such a new ``cosmological constant'' 
contribution to 
the entropy in the CV formula is quite remarkable. 

Thus far, 
we have discussed the appearance of the CV formula as a way to rewrite 
the FRW 
equations. However, the CV formula appears in a second 
formulation when one calculates
the entropy, $S$, of the universe. 
Indeed, following Ref. \cite{EV}, one can 
represent the total energy $E=\rho V$ of the universe 
as the sum of the 
extensive energy, $E_E$, and the subextensive (Casimir) 
energy $E_C$:
\be
\label{E10}
E(S,V)=E_E(S,V) + {1 \over 2}E_C(S,V)\ .
\ee
Note that unlike the case considered by Verlinde \cite{EV}, 
the cosmological 
constant contribution appears in $E_E$. 
Nevertheless, the constant rescaling of the energy is 
given by 
\bea
\label{F11}
E_E\left(\lambda S, \lambda V\right) &=& \lambda E_E
\left(S,V\right)\ ,\nn
E_C\left(\lambda S, \lambda V\right) &=& 
\lambda^{1-{2 \over n}}
E_C \left(S,V\right)\ .
\eea 

Now, if we assume that the first law of thermodynamics 
is valid and that the expansion is adiabatic, we deduce that 
\be
\label{F12}
dS=0\ ,\quad s={a^n \over T}\left(\rho + p\right) + s_0  ,
\ee
where the 
entropy $S \equiv s\int d^n x \sqrt{g}$, $s_0$ is an integration constant  
and $T$ is the 
temperature of the universe. It then follows that 
the Casimir energy is given by 
\cite{youm}
\be
\label{F13}
E_C=n\left(E + pV - TS\right) = -n Ts_0 \int d^nx \sqrt{g}  
\ee
and, consequently, that 
$E_C\sim a^{-n\omega}$ and $E_E - E_{\rm cosm}\sim a^{-n\omega}$. 
This further implies that the products $E_Ca^{n\omega}$ and 
$\left( E_E - E_{\rm cosm}\right)a^{n\omega}$ are independent of 
the spatial volume of the universe, $V$. 
By employing the scaling relations (\ref{F11}) one then concludes that 
\cite{youm}:
\be
\label{F14}
E_E - E_{\rm cosm}={\alpha \over 4\pi a^{n\omega}}S^{\omega+1}\ ,\quad 
E_C={\beta \over 2\pi a^{n\omega}}S^{\omega+1 - {2 \over n}}\ ,
\ee
where $\alpha$ and $\beta$ are arbitrary constants. 
Hence, the entropy is given by  
\be
\label{F15}
S=\left[{2\pi a^{n\omega} \over \sqrt{\alpha\beta}}
\sqrt{E_C\left(E_E - E_{\rm cosm}\right)} \right]^{n \over 
(\omega+1)n -1}\ .
\ee
Eq. (\ref{F15}) represents the generalization 
of the Cardy-Verlinde formula found by Youm \cite{youm} 
in the absence of a contribution from
the cosmological constant. The negative 
term associated with such a cosmological entropy is quite remarkable.
In the case of a radiation-dominated universe, 
Eq. (\ref{F15}) reduces to the standard CV formula
with the familiar square root term.

We may now proceed to 
consider whether a similar formulation of the 
Friedmann equation (\ref{RA5}) is possible 
for the Einstein--GB--Maxwell braneworld model. 
We may rewrite Eq. (\ref{RA5}) in the form:
\bea
\label{F1*}
H^2 &=& - {k \over 2a^2} + {\kappa_4^2 \over 6}
{\tilde E \over V}\ ,\nn
{\tilde E}&=&{6V_3 a^3 \over \kappa_4^2}
\left[\frac{{\cal{G}}^2}{{\cal{H}}^2} 
- {1 \over 2a^2} \left\{{a^2 \over 2\epsilon l^2}
\pm  {Y^{1/2} \over c}\right\}\right] \ ,\nn
V&=&a^3V_3\ ,
\eea
where $V_3$ is the volume of the three--dimensional sphere 
with  unit radius  and $\kappa_4$ is the four--dimensional 
gravitational coupling. 

In \cite{EV}, it was shown that the standard FRW equation in 
$d$ dimensions can be regarded as a $d$--dimensional analogue of 
the Cardy formula for a two--dimensional
CFT \cite{Cardy}:
\be
\label{CV1}
\tilde {\cal S}=2\pi \sqrt{
{c \over 6}\left(L_0 - {k \over d-2}{c \over 24}\right)}\ ,
\ee
where $c$ is the analogue of the two-dimensional central charge and 
$L_0$ is the analogue of the two-dimensional Hamiltonian. 
In the present case, making the identifications
\bea
\label{CV2}
{2\pi \tilde E r \over d-1} &\rightarrow& 2\pi L_0 \ ,\nn
{(d-2)V \over \kappa_d^2 r} &\rightarrow& {c \over 24} \ ,\nn
{4\pi (d-2)HV \over \kappa_d^2} &\rightarrow& \tilde {\cal S}\ ,
\eea
implies that 
the FRW-like equation (\ref{F1*}) has the same form as Eq. (\ref{CV1}). 
This is the first 
analogy that has been made with the CV formula, when one rewrites 
the FRW equation in a form similar to that of the Cardy formula.

It is also interesting to develop a relationship with thermodynamics.
In \cite{cvetic}, it was shown that 
the thermodynamical energy, $E$, and entropy, ${\cal S}$, of the bulk 
spacetime are given by 
\bea
\label{EnS2}
E&=&{3l^2 \over 16\kappa^2}V_3 \left(1 - 12\epsilon \right)
\left(k^2  + {16\kappa^4\tilde\mu \over l^4}\right) \ ,\\
\label{EnS4} 
{\cal S}&=&{V_3 \over \kappa^2}\left({1 - 12 \epsilon 
\over 1-4\epsilon} \right)\left(4\pi r_H^3 
+ 24 \epsilon k \pi r_H\right) + {\cal S}_0\ ,
\eea
where $r_H$ is the black hole radius, defined by the 
condition that  $\e^{2\nu}$ in Eq. 
(\ref{sol1}) vanishes at $r=r_H$. This 
constraint reduces to 
\be
\label{GBxxix}
0 = r_H^6 - {kl^2r_H^4 \over 2\left(2\epsilon -1 \right)} 
 - {\kappa^4\tilde\mu \left( 4\epsilon -1 \right)r_H^2 \over 
2\epsilon - 1 } 
- {\tilde Q^2 \kappa^4\over 2 \left(2\epsilon - 1\right)} 
\ee
and solving Eq. (\ref{GBxxix}) with respect to $\tilde\mu$, we 
find that 
\be
\label{EnS3}
\tilde\mu = {1 \over 2\kappa^4}
\left(k\epsilon - {1 \over 2}\right)^{-1}
\left\{ ( 2\epsilon - 1) r_H^4 
 - kr_H^2 l^2 - {\tilde Q^2\kappa^2 l^2 \over 2c r_H^2}\right\}\ .
\ee

It is unclear from the above relations 
how to relate the Friedmann equation to 
the thermodynamics of the bulk black hole
through a CV formula. In order to understand this issue more 
fully, therefore, 
we now consider a simpler 
five-dimensional theory that has a dual quantum field theory (QFT) analogue.
Specifically, instead of the Gauss-Bonnet theory, we can begin with 
the following five-dimensional, higher--order gravity theory:
\be
\label{ab1}
S=\int d^{d+1} x \sqrt{-\hat G}\left\{\hat a \hat R^2 
+ \hat b \hat R_{\mu\nu}\hat R^{\mu\nu}
+ {1 \over \kappa^2} \hat R - \Lambda \right\}\ ,
\ee
where $\{ \hat{a} , \hat{b} \}$ are 
coupling constants. 
In this case, the  Schwarzschild-anti de Sitter (SAdS) space 
is an exact solution \cite{NOO}:
\bea
\label{ab2}
ds^2&=&\hat G_{\mu\nu}dx^\mu dx^\nu \nn
&=&-\e^{2\rho_0}dt^2 + \e^{-2\rho_0}dr^2 
+ r^2\sum_{i,j}^{d-1} g_{ij}dx^i dx^j\ ,\nn
\e^{2\rho_0}&=&{1 \over r^{d-2}}\left(-\mu + {kr^{d-2} \over d-2} 
+ {r^d \over l^2}\right)\ ,
\eea
where $\mu$ parametrizes the mass 
of the black hole and the scale parameter, $l$, is determined by solving the 
constraint equation: 
\be
\label{abll}
0={80 \hat a \over l^4} + {16 \hat b \over l^4}
 - {12 \over \kappa^2 l^2}-\Lambda 
\ee
and the horizon radius, $r_{h}$, is deduced by solving 
the equation $e^{2\rho_0(r_H)}=0$ in (\ref{ab2}), i.e., 
\bea
\label{abrh1}
r_{H}^{2}=-{k l^{2} \over 4} + {1\over 2}
\sqrt{{k^2 \over 4}l^{4}+ 4\mu l^{2} } \; .
\eea
The Hawking temperature, $T_H$, is then given by 
\bea
\label{abht1}
T_H = {(e^{2\rho})'|_{r=r_{H}} \over 4\pi}
= {k \over 4\pi r_{H}} +{r_{H} \over \pi l^2}   ,
\eea
where a prime denotes differentiation with respect to $r$. 
One can also rewrite the mass parameter, $\mu$,  using $r_{H}$ or $T_{H}$
from Eq. (\ref{abrh1}) as follows:
\bea
\label{ab00}
\mu &=& {r_{H}^{4} \over l^{2}}+{kr_{H}^{2} \over 2}
=r_{H}^{2} \left( {r_{H}^{2} \over l^{2}}+{k \over 2} \right)\nn
&=& {1\over 4}\left( \pi l^{2} T_{H} \pm 
\sqrt{(\pi l^{2} T_{H})^{2} -kl^{2}} \right)^{2} \nn
&& \times \left({1\over 4 l^{2}}\left( \pi l^{2} T_{H} \pm
\sqrt{(\pi l^{2} T_{H})^{2} -kl^{2}} \right)^{2}
 +{k \over 2} \right)\ .
\eea 

The entropy ${\cal S}$ and the thermodynamical energy $E$ 
of the black hole are given by \cite{NOO}
\bea
\label{ab3}
{\cal S }&=& {V_{3}\pi r_H^3 \over 2}
\left( {8 \over \kappa^2}- {320 \hat a \over l^2}
 -{64 \hat b \over l^2} \right)\nn
&=& {V_{3}\pi \over 16} 
\left( \pi l^{2} T_{H} \pm 
\sqrt{(\pi l^{2} T_{H})^{2} -kl^{2}} \right)^3 
\left( {8 \over \kappa^2}- {320 \hat a \over l^2}
 -{64 \hat b \over l^2} \right)\ .\\
\label{ab4}
E&=& {3V_{3}\mu \over 8}
\left( {8 \over \kappa^2}- {320 \hat a \over l^2}
 -{64 \hat b \over l^2} \right)\ .
\eea
On the other hand, the FRW equations of the brane universe in the 
SAdS background are given by
\bea
\label{ab5}
&& H^2 = - {k \over 2a^2} + {\kappa_4^2 \over 6}
{\tilde E \over V}\ ,\quad
\dot H = - {\kappa_4^2 \over 4} \left({\tilde E \over V} 
+ p\right) + {k \over 2a^2}\ ,\nn
&& {\tilde E}={6 \mu V_3 \over \kappa_4^2 a}\ ,\quad
p={2\mu \over a^4 \kappa_4^2}\ ,\quad V=a^3V_3\ ,\nn
&& {1 \over \kappa_4^2} = {l \over 2}\left({1 \over \kappa^2}
 - {40 \hat a \over l^2} -{8 \hat b \over l^2} \right)\ .
\eea 
Since $3p={\tilde E}/V$, the trace of the 
energy-stress tensor arising from the matter on the brane 
vanishes, i.e.,  ${T^{{\rm matter}\ \mu}}_\mu=0$. 
Thus, the matter on the brane can be 
regarded as radiation or, equivalently, as massless fields. In other 
words, the field theory on the brane should be a conformal theory. 

We now assume that the total 
entropy ${\cal S}$ of the CFT on the brane is given 
by Eq. (\ref{ab3}).  
If this entropy is constant during the 
cosmological evolution, the entropy density $s$ is given by 
\be
\label{abe20}
s={{\cal S} \over a^3 V_3}
= {4\pi r_H^3 \over \tilde\kappa^2 a^3}
={8\pi r_H^3 \over l\kappa_4 a^3}\ .
\ee
If we further assume that the temperature $T$ on the brane 
differs from the Hawking temperature $T_H$ by the factor 
$l/a$ (for a discussion of why such a rescaling is necessary,
see 
\cite{SV}), it follows that 
\be
\label{abe22}
T={l \over a}T_H
={r_H \over \pi a l} + {kl \over 4\pi a r_H} 
\ee
and, when $a=r_H$, this implies that  
\be
\label{abe23}
T={1 \over \pi l} + {k \over 4\pi r_H^2}\ .
\ee
If the energy and entropy are purely extensive, the 
quantity $\tilde E + pV - T{\cal S}$ vanishes. In general, 
this condition does not hold and one can define the 
Casimir energy $E_C$ as in the above case for Einstein gravity: 
\be
\label{abEC1}
E_C=3\left(\tilde E + pV - T{\cal S}\right)\ .
\ee
Then, by using Eqs. (\ref{ab5}), (\ref{abe20}), and (\ref{abe22}), 
we find  that  
\be
\label{abEC2}
E_C={6k r_H^2 V \over \kappa_4^2 a^4} 
= {6k r_H^2 V_3 \over \kappa_4^2 a} 
= {3l k r_H^2 V_3 \over \kappa_4^2 a}\left(1
 - {40\hat a \kappa^2 \over l^2}
  - {8\hat b\kappa^2 \over l^2} \right) \ .
\ee
Consequently, the Casimir energy vanishes for $k=0$. 
When $\hat a$ and $\hat b$ are small, $E_C$ is positive (negative) for 
$k=2$ ($k=-2$). If, on the 
other hand, either $\hat a$ or $\hat b$ is large and positive, 
$E_C$ can be negative (positive) even when $k=2$ ($k=-2$).
Finally, by combining Eqs. (\ref{ab3}), (\ref{ab5}), and (\ref{abEC2}), 
for the case where $k\neq 0$, we find that
\be
\label{abSS}
{\cal S}={4\pi a \over 3\sqrt{|2k|}}\sqrt{\left| E_C\left(\tilde E 
 - E_C\right)\right|}\ .
\ee

Eq. (\ref{abSS}) has precisely the same form 
as the corresponding equation that arises in 
Einstein gravity. Thus, we have demonstrated how the 
FRW equation, when written in the CV form,  can be 
related to the thermodynamics of the bulk black hole.
However, in the theory where the GB term is present, 
it is anticipated that the dual QFT (if it exists) will not be a 
conformal theory. In this 
case, an extension of AdS/CFT correspondence to include
a non--CFT will be required and it 
is not clear how the FRW equation can be related to 
black hole thermodynamics in this scenario. 
It is possible that strong differences  
between the two approaches may arise. 

\section{Discussion}

\setcounter{equation}{0}

\def\theequation{\thesection.\arabic{equation}}

In this paper we have derived the brane 
Friedmann equation for the general, vacuum
FRW brane embedded in five--dimensional
Einstein--GB--Maxwell gravity, where the bulk spacetime 
may be interpreted as a charged black hole in an 
asymptotically AdS (or dS) space. 

Recently, brane dynamics of the second
Randall-Sundrum model \cite{RSII} has been discussed  
within the framework of Einstein--GB gravity 
\cite{AM,cs,CD}. Our approach is different to these previous works because 
we have considered  the surface terms that must be introduced into 
the bulk action (\ref{vi}) to 
ensure that the variational principle   
is well--defined. 
Moreover, our boundary action is fixed by the 
finiteness of the bulk space, or equivalently, by 
cancelling the leading--order divergences in the 
asymptotically AdS bulk. 

It is well known that in Einstein gravity, the variational principle is 
ill--defined when the manifold has a boundary, since the scalar curvature 
contains second--order derivatives of the metric, $g_{\mu\nu}$. 
Consequently, when the action is varied 
with respect to the metric, it acquires a non--trivial 
term that is proportional to the derivative of $\delta g_{\mu\nu}$ 
with respect to the coordinate that is 
transverse to the boundary. This term is cancelled by introducing the 
Gibbons--Hawking surface term \cite{GH} into the action. In this paper, 
the analogous counterterms  
for the GB contribution were introduced, together with the Gibbons--Hawking 
term and the leading counterterm arising from the brane vacuum 
energy. The crucial point is that once the action is varied, terms in 
$\delta A_{,y}$ no longer arise, thus ensuring  
that the variational principle is now
well--defined \cite{NOO}. Extremizing the action then led to the functional 
form of $A_{,y}$ and, hence, the effective
Friedmann equation (\ref{RA5}) on the brane. We emphasize that the 
introduction of the boundary 
terms is not {\em ad hoc}
and that their functional form is determined directly 
from the requirement that the variational 
principle be well defined and that the bulk space should be finite. 

When the higher--order
Gauss--Bonnet terms are present, 
the derivative terms in 
Eq. (\ref{viii})  
are non--trivial because the bulk spacetime 
is no longer an Einstein space. 
This leads to the extra complications 
in the form of $A_{,y}$, as summarized in Eq. (\ref{RA3}),
once the expressions (\ref{RA1}) and (\ref{RA2})
for the Ricci scalar and 
tensor are substituted. 

Consequently, the 
form of the Friedmann equation (\ref{RA5}) is itself 
highly non--trivial. Nevertheless, the qualitative dynamics can be 
established. The condition (\ref{cancel}) 
may be imposed on the parameters of the model by 
requiring the 
bulk action to be finite. When this condition is 
satisfied, the effective cosmological constant 
on the brane vanishes for the negative--branch cosmologies.
The main conclusion 
is that although the charge and mass of the bulk black hole 
are negligible at 
large spatial volumes, they radically alter the dynamics as the 
scale factor diminishes in size. 
In particular, 
the
Friedmann equation is dominated by a negative term in the 
limit of vanishing spatial volume, implying that such a limit 
is unphysical. 

Finally, we discussed two approaches where 
a (generalized) Cardy-Verlinde 
formula arises in FRW cosmology. The first approach rewrites
the FRW equations in a suitable form
and this method  may be adapted to the GB braneworld 
model considered in this paper.
In the second approach, 
the entropy of the universe is calculated in terms of a dual 
quantum field theory.
However, at present, it is not clear how this technique may be employed within 
the context of the GB braneworld scenario, since it is expected that the
QFT that is dual to the AdS black hole should be non-conformal
in the presence of 
a GB 
term.

\vspace{.3in}

\centerline{\bf Acknowledgments}

\vspace{.3in}

JEL is supported by the Royal Society.  
The work by SN is supported in part by the Ministry of Education, 
Science, Sports and Culture of Japan under the grant number 13135208.

\newpage

\appendix

\section{Appendix}

\def\theequation{A.\arabic{equation}}

\setcounter{equation}{0}

\def\theequation{\thesection.\arabic{equation}}

When the five--dimensional metric has the form given in 
Eqs. (\ref{metric1}) and (\ref{worldvol}), the non--vanishing 
components of the curvature tensor 
and its contractions are:
\bea
\label{crvtrs}
 R_{yiyj}&=&\e^{2A}\left(-A_{,yy} -
\left(A_{,y}\right)^2\right)\tilde g_{ij} \nn
 R_{yA\sigma B}&=&-l^2\e^{2A}A_{,y\sigma}g^s_{AB} \nn
 R_{\sigma A\sigma B}&=&\left(-l^2 \e^{2A}A_{,\sigma\sigma}
-l^2 \e^{4A}\left(A_{,y}\right)^2\right)g^s_{AB} \nn
 R_{ABCD}&=&\left(l^2 \e^{2A} -l^2 \e^{2A}
\left(A_{,\sigma}\right)^2
- l^4\e^{4A}\left(A_{,y}\right)^2\right)
\left(g^s_{AC}g^s_{BD} - g^s_{AD}g^s_{BC}\right) \nn
 R_{yy}&=&4\left(-A_{,yy} - \left(A_{,y}\right)^2\right) \nn
 R_{y\sigma}&=&-3A_{,y\sigma} \nn
 R_{\sigma\sigma}&=&l^2\e^{2A}\left(-A_{,yy}
- 4\left(A_{,y}\right)^2\right) - 3A_{,\sigma\sigma} \nn
 R_{AB}&=&\left( l^2\e^{2A}\left(-A_{,yy}
- 4\left(A_{,y}\right)^2\right) - A_{,\sigma\sigma}
- 2\left(A_{,\sigma}\right)^2 +2\right) g^s_{AB} \nn
 R&=&-8A_{,yy} - 20\left(A_{,y}\right)^2 +l^{-2}\e^{-2A}
\left(-6A_{,\sigma\sigma} - 6\left(A_{,\sigma}\right)^2 +6
\right) ,
\eea
where $g^s_{AB}$ represents the metric on the three--space 
with the line--element given by $d\Omega^2_{k,3}$ in Eq. 
(\ref{worldvol}).


\begin{thebibliography}{99}

\bibitem{adscft}
J. M. Maldacena, 
``The Large N Limit of Superconformal Field Theories and Supergravity'',
{\em Adv. Theor. Math. Phys.} {\bf 2} (1998) 231, \\
{\tt [http://arXiv.org/abs/hep-th/9711200]}. \\

E. Witten, 
``Anti De Sitter Space And Holography'',
{\em Adv. Theor. Math. Phys.} {\bf 2} (1998) 253, \\
{\tt [http://arXiv.org/abs/hep-th/9802150]}. \\

S. Gubser, I. Klebanov, and A. Polyakov, 
``Gauge Theory Correlators from Non-Critical String Theory'',
{\em Phys. Lett.} {\bf B428} (1998) 105, \\
{\tt [http://arXiv.org/abs/hep-th/9802109]}. \\

O. Aharony, S. Gubser, J. Maldacena, H. Ooguri, and Y. Oz,
``Large N Field Theories, String Theory and Gravity'',
{\it Phys. Rep.} {\bf 323} (2000) 183, \\ 
{\tt [http://arXiv.org/abs/hep-th/9905111]}. 


\bibitem{dscft} 
A. Strominger, 
``The dS/CFT Correspondence'',
{\em JHEP} {\bf 10} (2001) 034, \\
{\tt [http://arXiv.org/abs/hep-th/0106113]}. \\

M. Spardlin, A. Strominger, and A. Volovich, 
``Les Houches Lectures on De Sitter Space'',
{\tt [http://arXiv.org/abs/hep-th/0110007]}.

\bibitem{hull}
C. M. Hull, 
``Timelike T-Duality, de Sitter Space, 
Large $N$ Gauge Theories and Topological Field Theory'',
{\em JHEP} {\bf 07} (1998) 021, \\
{\tt [http://arXiv.org/abs/hep-th/9806146]}.

\bibitem{branerefs}
K. Akama, 
``An Early Proposal of Brane World'',
{\em Lect. Notes Phys.} {\bf 176} (1982) 267, \\
{\tt [http://arXiv.org/abs/hep-th/0001113]}. \\

V. A. Rubakov and M. E. Shaposhnikov, 
``Extra Spacetime Dimensions: Towards a Solution to the Cosmological 
Constant Problem'', 
{\em Phys. Lett.} {\bf 125B} (1983) 136. \\

M. Visser, 
``An Exotic Class of Kaluza--Klein Models'', 
{\em Phys. Lett.} {\bf B159} (1985) 22. \\

N. Arkani--Hamed, S. Dimopoulos, and G. Dvali, 
``The Hierarchy Problem and New Dimensions at a Millimeter'',
{\em Phys. Lett.} {\bf B429} (1998) 263, \\
{\tt [http://arXiv.org/abs/hep-ph/9803315]}. \\

I. Antoniadis, N. Arkani--Hamed, S. Dimopoulos, and G. Dvali, 
``New Dimensions at a Millimeter to a Fermi and Superstrings at a TeV'',
{\em Phys. Lett.} {\bf B436} (1998) 257, \\
{\tt [http://arXiv.org/abs/hep-ph/9804398]}. \\

M. Gogberashvili, 
``Our World as an Expanding Shell'',
{\em Europhys. Lett.} {\bf 49} (2000) 396, \\
{\tt [http://arXiv.org/abs/hep-ph/9812365]}. \\

L. Randall and R. Sundrum, 
``A Large Mass Hierarchy from a Small Extra Dimension'',
{\em Phys. Rev. Lett.} {\bf 83} (1999) 3370, \\
{\tt [http://arXiv.org/abs/hep-th/9905221]}.  

\bibitem{RSII} L. Randall and R. Sundrum, 
``An Alternative to Compactification'',
{\em Phys. Rev. Lett.} {\bf 83} (1999) 4690, \\
{\tt [http://arXiv.org/abs/hep-th/9906064]}.  


\bibitem{othermoves}
P. Kraus, 
``Dynamics of Anti-de Sitter Domain Walls'',
{\em JHEP} {\bf 12} (1999) 011, \\
{\tt [http://arXiv.org/abs/hep-th/9910149]}. \\

H. A. Chamblin and H. S. Reall, 
``Dynamic Dilatonic Domain Walls'',
{\em Nucl. Phys.} {\bf B562} (1999) 133, \\
{\tt [http://arXiv.org/abs/hep-th/9903225]}. \\


H. A. Chamblin, M. J. Perry, and H. S. Reall, 
``Non-BPS D8-branes and Dynamic Domain Walls in Massive 
IIA Supergravities'',
{\em JHEP} {\bf 09} (1999) 014, \\
{\tt [http://arXiv.org/abs/hep-th/9908047]}. \\

H. Collins and B. Holdom, 
``Brane Cosmologies without Orbifolds'',
{\em Phys. Rev.} {\bf D62} (2000) 105009, \\
{\tt [http://arXiv.org/abs/hep-ph/0003173]}. \\

S. Nojiri and S. D. Odintsov, 
``Brane World Inflation Induced by Quantum Effects'',
{\em Phys. Lett.} {\bf B484} (2000) 119, \\
{\tt [http://arXiv.org/abs/hep-th/0004097]}.  

\bibitem{bv}
C. Barcelo and M. Visser, 
``Living on the edge: cosmology on the boundary of anti-de Sitter space'',
{\em Phys. Lett.} {\bf B482} (2000) 183, \\
{\tt [http://arXiv.org/abs/hep-th/0004056]}.  

\bibitem{ida}
D. Ida, 
``Brane-world cosmology'',
{\em JHEP} {\bf 09} (2000) 014, \\
{\tt [http://arXiv.org/abs/gr-qc/9912002]}. 


\bibitem{bine}
P. Bin\'etruy, C. Deffayet, U. Ellwanger, and D. Langlois, 
``Brane cosmological evolution in a bulk with cosmological constant'',
{\em Phys. Lett.} {\bf B477} (2000) 285, \\
{\tt [http://arXiv.org/abs/hep-th/9910219]}. \\

E. E. Flanagan, S. -H. Tye, and I. Wasserman, 
``Cosmological Expansion in the Randall-Sundrum Brane World Scenario'',
{\em Phys. Rev.} {\bf D62} (2000) 044039, \\
{\tt [http://arXiv.org/abs/hep-ph/9910498]}. \\

S. Mukohyama, 
`` Brane-world solutions, standard cosmology, and dark radiation'',
{\em Phys. Lett.} {\bf B473} (2000) 241, \\
{\tt [http://arXiv.org/abs/hep-th/9911165]}.  

\bibitem{maeda}
K. Maeda, 
``Brane Quintessence'', 
{\em Phys. Rev.} {\bf D64} (2001) 123525, \\
{\tt [http://arXiv.org/abs/astro-ph/0012313]}. \\

S. Mizuno and K. Maeda, 
``Quintessence in a Brane World'',
{\em Phys. Rev.} {\bf D64} (2001) 123521, \\
{\tt [http://arXiv.org/abs/hep-ph/0108012]}. \\

E. J. Copeland, A. R. Liddle, and J. E. Lidsey, 
``Steep inflation: ending braneworld inflation by 
gravitational particle production'',
{\em Phys. Rev.} {\bf D64} (2001) 023509, \\
{\tt [http://arXiv.org/abs/astro-ph/0006421]}. \\

R. M. Hawkins and J. E. Lidsey, 
``Inflation on a single brane - exact solutions'', 
{\em Phys. Rev.} {\bf D63} (2001) 041301, \\
{\tt [http://arXiv.org/abs/gr-qc/0011060]}. \\


G. Huey and J. E. Lidsey, 
`` Inflation, braneworlds and quintessence'',
{\em Phys. Lett.} {\bf B514} (2001) 217, \\
{\tt [http://arXiv.org/abs/astro-ph/0104006]}. \\

R. M. Hawkins and J. E. Lidsey, 
``The Ermakov-Pinney Equation in Scalar Field Cosmologies'', \\
{\tt [http://arXiv.org/abs/astro-ph/0112139]}. 

\bibitem{HD}
S. Nojiri and S. D. Odintsov, 
`Can We Live on the Brane in Schwarzschild-anti 
de Sitter Black Hole?'', \\
{\em Phys. Lett.} {\bf B493} (2000) 153, \\
{\tt [http://arXiv.org/abs/hep-th/0007205 ]}.


S. Nojiri and S. D. Odintsov, 
``Brane-World Cosmology in Higher Derivative Gravity or 
Warped Compactification in the Next-to-leading Order of
AdS/CFT Correspondence'',
{\em JHEP} 07 (2000) 049, \\
{\tt [http://arXiv.org/abs/hep-th/0006232]}. \\

M. Fukuma, S. Matsuura, 
``Holographic Renormalization Group Structure 
in Higher-Derivative Gravity'', 
{\tt [http://arXiv.org/abs/hep-th/0112037]}. \\


M. Giovannini,
``Static dilaton solutions and singularities in six dimensional warped 
compactification with higher derivative'',
{\em Phys. Rev.} {\bf D63} (2001) 064011, \\
{\tt [http://arXiv.org/abs/hep-th/0111153]}. \\


S. Mukohyama, 
`` Static solutions in the $R^4$ brane world'',
{\em Phys. Rev.} {\bf D63} (2001) 104025, \\
{\tt [http://arXiv.org/abs/hep-th/0101038]}. \\


G. Kofinas, 
``General Brane Dynamics with $R^{4}$ term in the Bulk''
{\em JHEP} {\bf 08} (2001) 034, \\
{\tt [http://arXiv.org/abs/hep-th/0108013]}. \\

\bibitem{NOO} S. Nojiri, S. D. Odintsov, and S. Ogushi,
``Holographic entropy and brane FRW-dynamics from AdS black hole 
in 5d higher derivative gravity'',
{\em Int. J. Mod. Phys.} {\bf A16} (2001) 5085, \\
{\tt [http://arXiv.org/abs/hep-th/0105117]}. \\

S. Nojiri, S. D. Odintsov, and S. Ogushi,
``Cosmological and black hole brane-world universes in higher derivative 
gravity'', 
{\em Phys. Rev.} {\bf D65} (2002) 023521, \\
{\tt [http://arXiv.org/abs/hep-th/0108172]}. 


\bibitem{GBrefs}
J. E. Kim, B. Kyae, and H. M. Lee, 
``Effective Gauss-Bonnet Interaction in Randall-Sundrum Compactification'',
{\em Phys. Rev.} {\bf D62} (2000) 045013, \\
{\tt [http://arXiv.org/abs/hep-ph/9912344]}. \\

J. E. Kim and H. E. Lee, 
``Gravity in the Einstein-Gauss-Bonnet Theory with the 
Randall-Sundrum Background'',
{\em Nucl. Phys.} {\bf B602} (2001) 346, Erratum--{\em ibid}
{\bf B619} (2001) 763,  \\
{\tt [http://arXiv.org/abs/hep-th/0010093]}. \\

N. Deruelle and T. Dolezel, 
``Brane versus shell cosmologies in Einstein and 
Einstein-Gauss-Bonnet theories'',
{\em Phys. Rev.} {\bf D62} (2000) 103502, \\
{\tt [http://arXiv.org/abs/gr-qc/0004021]}. \\

N. E. Mavromatos and J. Rizos, 
``String-Inspired Higher-Curvature Terms and the Randall-Sundrum Scenario'',
{\em Phys. Rev.} {\bf D62} (2000) 124004, \\
{\tt [http://arXiv.org/abs/gr-qc/0008074]}. \\

I. Low and A. Zee, 
``Naked Singularity and Gauss-Bonnet Term in Brane World Scenarios'',
{\em Nucl. Phys.} {\bf B585} (2000) 395, \\
{\tt [http://arXiv.org/abs/hep-th/0004124]}. \\

O. Corradini and Z. Kakushadze, 
``Localized Gravity and Higher Curvature Terms'',
{\em Phys. Lett.} {\bf B494} (2000) 302, \\
{\tt [http://arXiv.org/abs/hep-th/0009022]}. \\

K. Meissner and M. Olechowski, 
``Domain walls without cosmological constant in higher order gravity'', 
{\em Phys. Rev. Lett.} {\bf 86} (2001) 3708, \\
{\tt [http://arXiv.org/abs/hep-th/0009122]}. \\

K. Kashima, 
``The M2-brane Solution of Heterotic M-theory with the Gauss-Bonnet 
$R^2$ terms'', 
{\em Prog. Theor. Phys.} {\bf 105} (2001) 301, \\
{\tt [http://arXiv.org/abs/hep-th/0010286]}. \\

Y. M. Cho and  I. P. Neupane, 
``Anti-de Sitter Black Holes, Thermal Phase Transition and Holography 
in Higher Curvature Gravity'', \\
{\tt [http://arXiv.org/abs/hep-th/0202140]}.  

\bibitem{onlylocal}
I. P. Neupane, 
``Gravitational potential correction with Gauss-Bonnet interaction'',
{\em Phys. Lett.} {\bf B512} (2001) 137, \\
{\tt [http://arXiv.org/abs/hep-th/0104226]}. \\


Y. M. Cho, I. Neupane, and P. S. Wesson, 
``No ghost state of Gauss-Bonnet interaction in warped backgrounds'',
{\em Nucl. Phys.} {\bf B621} (2002) 388, \\
{\tt [http://arXiv.org/abs/hep-th/0104227]}. \\

J. E. Kim, B. Kyae, and H. M. Lee, 
``Various Modified Solutions of the Randall-Sundrum Model 
with the Gauss-Bonnet Interaction'',
{\em Nucl. Phys.} {\bf B582} (2000) 296; 
Erratum--{\em ibid} {\bf B591}  (2000) 587,  \\
{\tt [http://arXiv.org/abs/hep-th/0004005]}.



\bibitem{AM} 
B. Abdesselam and N. Mohammedi, 
``Brane World Cosmology with Gauss-Bonnet Interaction'', \\
{\tt [http://arXiv.org/abs/hep-th/0110143]}. 


\bibitem{cs} 
C. Germani and C. F. Sopuerta, 
``String inspired braneworld cosmology'', \\
{\tt [http://arXiv.org/abs/hep-th/0202060]}.

\bibitem{CD} 
C. Charmousis and J. Dufaux, 
``General Gauss-Bonnet brane cosmology'', \\
{\tt [http://arXiv.org/abs/hep-th/0202107]}.

\bibitem{largeN}
A. Fayyazuddin and M. Spalinski, 
``Large N Superconformal Gauge Theories and Supergravity Orientifolds'',
{\em Nucl. Phys.} {\bf B535} (1998) 219, \\
{\tt [http://arXiv.org/abs/hep-th/9805096]}. \\

O. Aharony, A. Fayyazuddin, and J. Maldacena, 
``The Large N Limit of ${\cal N} =2,1 $ Field Theories 
from Threebranes in F-theory'',
{\em JHEP} {\bf 07} (1998) 013, \\
{\tt [http://arXiv.org/abs/hep-th/9806159]}.

\bibitem{stringGB}
B. Zwiebach, 
``Curvature Squared Terms and String Theories'',
{\em Phys. Lett.} {\bf 156B} (1985) 315. \\

A. Sen, 
``Equations of Motion for the Heterotic String Theory from the 
Conformal Invariance of the Sigma Model'',
{\em Phys. Rev. Lett.} {\bf 55} (1985) 1846. \\

R. R. Metsaev and A. A. Tseytlin, 
``Order $\alpha'$ (Two--Loop) Equivalence of the String Equations 
of Motion and the $\sigma$--model Weyl Invariance Conditions'',
{\em Nucl. Phys.} {\bf B293} (1987) 385. 

\bibitem{bd}
D. G. Boulware and S. Deser, 
``String--Generated Gravity Models'', \\
{\em Phys. Rev. Lett.} {\bf 55} (1985)  2656. 

\bibitem{d86}
N. Deruelle and J. Madore, 
``The Friedmann Universe as an Attractor in a Kaluza--Klein 
Cosmology'',
{\em Mod. Phys. Lett.} {\bf A1} (1986) 237. \\

N. Deruelle and L. Farina--Busto, 
``Lovelock Gravitational Field Equations in Cosmology'',
{\em Phys. Rev.} {\bf D41} (1990) 3696.

\bibitem{cai} 
R.-G. Cai,
``Gauss-Bonnet Black Holes in AdS Spaces'', \\
{\tt [http://arXiv.org/abs/hep-th/0109133]}.

\bibitem{cvetic} M. Cvetic, S. Nojiri, and S. D. Odintsov,
``Black Hole Thermodynamics and Negative Entropy in de Sitter and 
Anti-de Sitter Einstein-Gauss-Bonnet Gravity'', \\
{\tt [http://arXiv.org/abs/hep-th/0112045]}.

\bibitem{EV} E. Verlinde, 
``On the Holographic Principle in a Radiation Dominated Universe'', \\
{\tt [http://arXiv.org/abs/hep-th/0008140]}.

\bibitem{Cardy} 
J. L. Cardy, 
``Operator Content of Two--dimensional Conformally Invariant Theories'',
{\em Nucl. Phys.} {\bf B270} (1986) 186.

\bibitem{hawkingpage}
S. W. Hawking and D. N. Page, 
``Thermodynamics of Black Holes In Anti--de Sitter Space''
{\em Commun. Math. Phys.} {\bf 87} (1983) 577. 

\bibitem{GH} G. W. Gibbons and S. W. Hawking, 
``Action Integrals and Partition Functions in Quantum Gravity'',
{\em Phys. Rev.} {\bf D15} (1977) 2752. 

\bibitem{csaki}
G. Csaki, J. Erlich, and C. Grojean, 
``Gravitational Lorentz Violations and Adjustment of the 
Cosmological Constant in Asymmetrically Warped Spacetimes'',
{\em Nucl. Phys.} {\bf B604} (2001) 312, \\
{\tt [http://arXiv.org/abs/hep-th/0012143]}.

\bibitem{youm} D. Youm, 
``A Note on the Cardy-Verlinde Formula'',
{\tt [http://arXiv.org/abs/hep-th/0201268]}.

\bibitem{wang} B. Wang, E. Abdalla, and R. Su, 
``Friedmann equation and Cardy formula correspondence in brane universes'',
{\em Phys. Lett.} {\bf B503} (2001) 394, \\
{\tt [http://arXiv.org/abs/hep-th/0106086]}.

\bibitem{ryan} 
S. Nojiri, O. Obregon, S. D. Odintsov, H. Quevedo, and M. P. Ryan,  
`` Quantum bounds for gravitational de Sitter entropy and the 
Cardy-Verlinde formula'',
{\em Mod. Phys. Lett.} {\bf A16} (2001) 1181, \\
{\tt [http://arXiv.org/abs/hep-th/0105052]}. 

\bibitem{SV} 
I. Savonije and E. Verlinde, 
``CFT and Entropy on the Brane'', 
{\em Phys. Lett.} {\bf B507} (2001) 305, \\
{\tt [http://arXiv.org/abs/hep-th/0102042]}. 




\end{thebibliography}
\end{document}